**Molybdenum Trioxide Gates for Suppression of Leakage Current in InAlN/GaN HEMTs at 300°C**

*Caitlin A. Chapin\*, Savannah R. Benbrook\*, Chloe Leblanc, Debbie G. Senesky\**

Dr. C. A. Chapin, Prof. D. G. Senesky
Aeronautics and Astronautics Department
Stanford University
Durand Building
496 Lomita Mall, Stanford, CA 94305, USA
E-mail: cchapin3@stanford.edu; dsenesky@stanford.edu

S. R. Benbrook, C. Leblanc
Electrical Engineering Department
Stanford University
David Packard Building
350 Serra Mall, Stanford, CA 94305, USA
E-mail: sbenbroo@stanford.edu

Because high electron mobility transistors (HEMTs) often exhibit significant gate leakage during high-temperature operation, the choice of Schottky metal is critical. Increased gate leakage and reduced ON/OFF ratio are unsuitable for the design of high-temperature electronics and integrated circuits. This paper presents high-temperature characteristics of depletion-mode molybdenum trioxide ($MoO_3$)-gated InAlN/GaN-on-silicon HEMTs in air. After a room temperature oxidation of the Mo for 10 weeks, the leakage of the HEMT is reduced over 60 times compared to the as-deposited Mo. The use of $MoO_3$ as the Schottky gate material enables low gate leakage, resulting in a high ON/OFF current ratio of $1.2\times10^8$ at 25°C and $1.2\times10^5$ at 300°C in air. At 400°C, gate control of the InAlN/GaN two-dimensional electron gas (2DEG) channel is lost and unrecoverable. Here, this permanent device failure is attributed to volatilization of the $MoO_3$ gate due to the presence of water vapor in air. Passivation of the device with SiN enables operation up to 500°C, but also increases the leakage current. The suppression of gate leakage via Mo oxidation and resulting high

ON/OFF ratio paves the way for viable high-temperature GaN-based electronics that can function beyond the thermal limit of silicon once proper passivation is achieved.

## 1. Introduction

Electronics that can survive in oxidizing, high-temperature environments are increasingly required by the space, military, automotive, and energy industries. Wide-bandgap semiconductors, such as gallium nitride (GaN), have emerged as promising extreme environment material platforms, due to their low intrinsic carrier concentrations (~$10^{-10}$ cm$^{-3}$ at room temperature (RT)) and high Schottky barrier heights (0.5-1.5 eV).[1, 2] GaN high electron mobility transistors (HEMTs) rely on the 2-dimensional electron gas (2DEG) conducting channel that occurs at the interface of GaN and III-nitride alloys (e.g. InAlN and AlGaN). While the stability of the GaN-heterostructure has been demonstrated up to 1000°C in vacuum, device failure is often reported to occur at much lower temperatures due to degradation mechanisms associated with the ohmic and Schottky contacts and passivation schemes.[3–7] The reduction in gate leakage is critical for high-temperature operation because it is the major source of OFF-current in depletion-mode HEMTs and leakage mechanisms (e.g., thermionic emission, thermionic trap-assisted tunneling) are generally inversely proportional to $e^{1/kT}$.[8–12]

Multiple approaches have been explored to develop GaN-heterostructure HEMTs with high ON/OFF ratios and low leakage currents including: introducing gate dielectrics for metal insulator semiconductor (MIS) structures (e.g., Hf$_2$O, TaO$_x$N$_y$), surface treatments prior to gate deposition (e.g., O$_2$ plasma + HCl, SF$_6$ plasma, KOH), post evaporation gate anneals (e.g. N$_2$/H$_2$, N$_2$), backfill of mesa isolation, use of unannealed ohmic contacts to avoid trap introduction, and the use of conductive refractory metal oxides (e.g., Ir/Al oxide).[13-20] While enhancement-mode InAlN/GaN devices have reached ON/OFF ratios of $10^{12}$ [21], RT ON/OFF ratios for depletion-mode (d-mode) ON/OFF ratios for InAlN HEMTs are reported in the range $10^3$-$10^7$, with Herfurth et al. achieving the only $10^{10}$ performance [17, 22-24]. Chen et al.

demonstrated an Al$_2$O$_3$ passivated d-mode InAlN/GaN-on-Si HEMT utilizing an N$_2$ annealed Ni Schottky gate with RT ON/OFF ratio of 10$^7$ and Ganguly et al. reported a d-mode InAlN/GaN-on-SiC HEMT using a KOH surface treatment and a Ni/Au gate with RT ON/OFF ratio of 10$^6$. [17, 23]

There have been several compelling high-temperature studies of InAlN/GaN HEMTs in inert ambient or in vacuum utilizing different Schottky gate and dielectric stacks. Herfurth et al. demonstrated a high ON/OFF ratio of 10$^{10}$ at RT and 10$^6$ at 600°C in vacuum in an ultra-thin body InAlN/GaN HEMT with Cu/Pt gate metal deposited onto a 1.5nm native oxide on the surface. [24] In the most extreme test of the InAlN/GaN HEMT to date, Maier *et al.* demonstrated a lattice-matched InAlN/GaN HEMT with a molybdenum gate metal and SiN passivation for 25 hours in vacuum at 1000°C. [3] However, ON/OFF ratio is not reported. There have been no studies on high-temperature (>300°C) operation of depletion-mode InAlN/GaN HEMTs in air. It is necessary to investigate the effects of the air environment on high-temperature performance of these devices because not all applications permit hermetic sealing to protect from the environment. The effects of oxidation on the contact metallization, barrier layer surface, and 2DEG charge density must be further studied for the InAlN/GaN material platform.

In this work, we present the first results of the impact of molybdenum trioxide (MoO$_3$) gate metallization on leakage current and ON/OFF ratio for an unpassivated depletion-mode InAlN/GaN HEMT up to 300°C in air. After oxidation in air for 10 weeks the leakage current decrease by ~60 times. The device demonstrates a high ON/OFF drain current ratio of 1.2×10$^8$ at 25°C and 1.2×10$^5$ at 300°C in air. However, loss of gate control and subsequent device failure at 400°C was found to be caused by volatilization of the MoO$_3$ in air.

2. Experimental Section

2.1 Device Fabrication

The HEMT devices were fabricated with an In$_{0.18}$Al$_{0.82}$N/GaN-on-silicon wafer purchased from NTT Advanced Technology Corporation. The GaN heterostructure, schematic of the gate evolution, and an optical image of the device is shown in **Figure 1**. Devices were isolated through $BCl_3/Cl_2$ ICP mesa etching, followed by evaporation of Ti/Al/Mo/Au (10/200/40/80 nm) ohmic contacts which were annealed at 850°C for 35 seconds in $N_2$. Molybdenum (30 nm) was evaporated as the Schottky contact and Ti/Au bond metal was patterned with liftoff. The Mo gates were left to oxidize at RT in air to form $MoO_3$ for ten weeks. The gate length of the HEMT is 1 µm, centered between source and drain (low power) in a 5-µm-long by 50-µm-wide 2DEG channel.

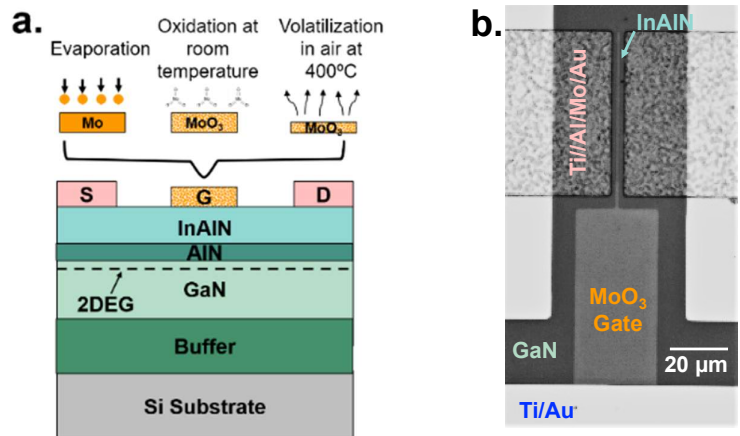

**Figure 1.** Schematic of InAlN/GaN heterostructure and molybdenum gate evolution due to high-temperature exposure in air (a) and optical image of microfabricated HEMT transistor (b).

## 2.2 Experimental Setup

The high-temperature response of the InAlN/GaN HEMTs was characterized on a Signatone Inc. high-temperature probe station from RT to 300°C in 50°C increments in air, with a holding time of ten minutes at each temperature before data was acquired. To examine the return characteristics, measurements were then taken in 100°C increments as the temperature was ramped back down. Finally, the device was ramped directly up to 400°C and tested. Current-voltage ($I_D$-$V_{DS}$, $I_D$-$V_{GS}$, $I_G$-$V_{GS}$) measurements were taken on a semiconductor parameter analyzer (Agilent Technologies B1500A).

## 3. Results and Discussion

**Figure 2**(a) compares the $I_D$-$V_{GS}$ curves of the InAlN HEMT at 10 mV drain to source voltage one week after Mo deposition and 10 weeks after the deposition. The long duration between the first test and the second test enabled the $MoO_3$ to undergo a RT oxidation to $MoO_3$. The off current of the device drops by ~60 times after the RT anneal, showing the oxidation process reduces leakage in the off state.

The high temperature response, $I_D$-$V_{DS}$, $I_D$-$V_{GS}$, and $I_G$-$V_{GS}$, after undergoing the 10-week RT anneal are shown in Figure 2(b) and (c). The large gate leakage is the dominate form of OFF-current when the device is operated below the threshold voltage (Figure 2(b)). The $I_D$-$V_{DS}$ response shows the performance when ramping up to 300°C and then back down to 25°C. Solid lines represent measurements taken as the temperature was increased and dotted lines represent the measurements taken as the temperature is decreased. The characteristics show little difference in performance during temperature ramp up and ramp down.

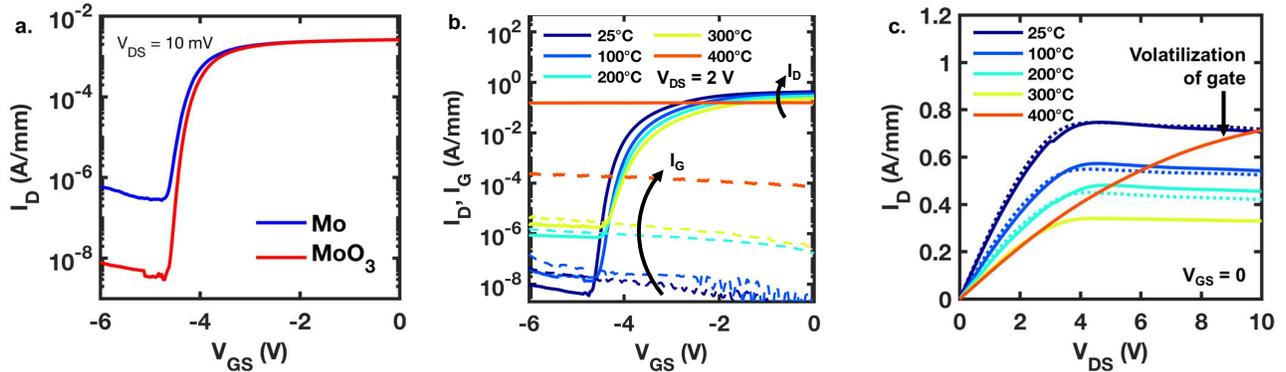

**Figure 2.** $I_D$-$V_{GS}$ curves comparing as deposited Mo and $MoO_3$ (a). $I_D$-$V_{GS}$ (solid) and $I_G$-$V_{GS}$ (dashed) curves from RT to 400°C (b). $I_D$-$V_{DS}$ curves from RT to 400°C; dotted lines represent the return response while ramping down to RT from 300°C (c).

After testing to 300°C and performing elemental characterization, the device was brought directly up to 400°C. While the device showed promising performance up to 300°C, testing at 400°C caused non-recoverable device failure, due to loss of gate control, shown in Figure 2(b) and 2(c). After this failure at 400°C, the gate is no longer visible under an optical microscope. Auger electron spectroscopy (AES) is used to confirm the disappearance of the gate after exposure to 400°C. AES maps of the device after 300°C and 400°C exposure are shown in **Figure 3**(a) and (b). The AES data shows that the Mo elemental ratio goes from ~34% to ~6% after 300°C and 400°C exposure, respectively.

X-Ray photoluminescence spectroscopy (XPS) measurements were taken to determine whether the Mo had remained a metal or had oxidized. Figure 3(c) shows the XPS spectra of the Mo $3d_{3/2}$ and $3d_{5/2}$ peaks after a 10-week RT anneal and after 300°C exposure. Fits find that the Mo $3d_{3/2}$ and $3d_{5/2}$ peaks are at 232.7 eV and 235.9 eV with no exposure to temperature and at 232.4 eV and 235.5 eV after 300°C exposure. Both of these sets of peaks match previously reported binding energies for the 3d orbital for $MoO_3$.[25, 26] This shows that the molybdenum oxidizes even under RT conditions.

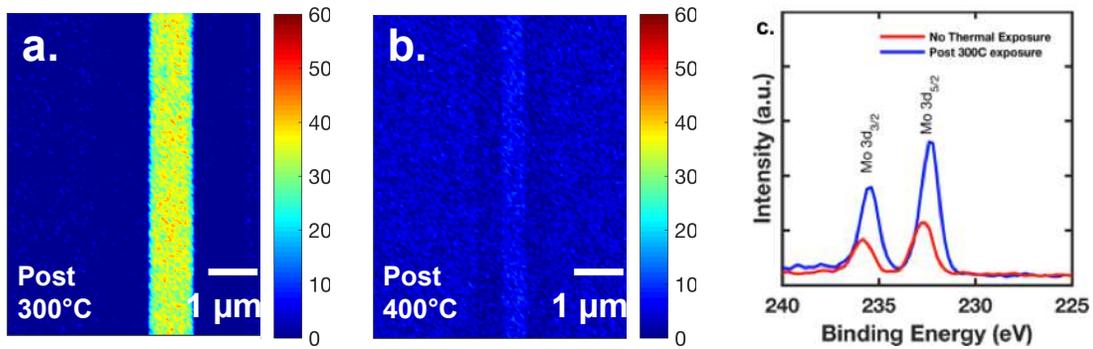

**Figure 3.** AES map of percent molybdenum after 300°C exposure (a) and after 400°C exposure (b). XPS plot of binding energy versus intensity of Mo $3d_{3/2}$ and $3d_{5/2}$ peaks (c).

Previous work has shown that $MoO_3$ volatizes when exposed to water vapor at elevated temperatures to form $MO_2(OH)_2$:[27, 28]

$$MoO_3 \text{ (s)} + H_2O \text{ (g)} \rightarrow MO_2(OH)_2 \text{ (g)} \qquad (1).$$

Thus, it is hypothesized that at 400°C the $MoO_3$ volatilizes in air at a rate significant enough to cause loss of the Schottky metal, likely due to reaction with water vapor. Temperatures under 400°C should not be thought of as safe because volatilization rates are exponentially temperature dependent. [27] Thus, the 300°C operation would not be exempt from loss of the gate, but the volatilization occurs less quickly.

While the unpassivated MoO$_3$ system is fragile and cannot exceed temperatures of 400°C, the initial device performance below 300°C is promising and surpasses the operational limits of silicon. The off current ($I_{D,Off}$), the saturation current ($I_{D,sat}$) at $V_{DS}$ = 5 V and $V_{GS}$ = 0 V, and the ON/OFF ratio ($I_{D,On}/I_{D,Off}$) of the device before failure are shown in **Figure 4**. Red markers and blue markers represent measurements taken as the temperature ramped up and ramped down, respectively. The ON/OFF ratio of the device is 1.2×10$^8$ and 1.2×10$^5$ at 25°C and 300°C, respectively. While the saturation current does not show degreadation after 300°C exposure, the ON/OFF ratio decreases to 2.9×10$^7$ after returning to 25°C, which can be attributed to the increase in the OFF-state current. In Figure 4, the leakage current is increasing exponentially with increasing temperature and is attributed to thermionic emission, illustrating the challenge of maintaining large ON/OFF ratios at high temperatures.

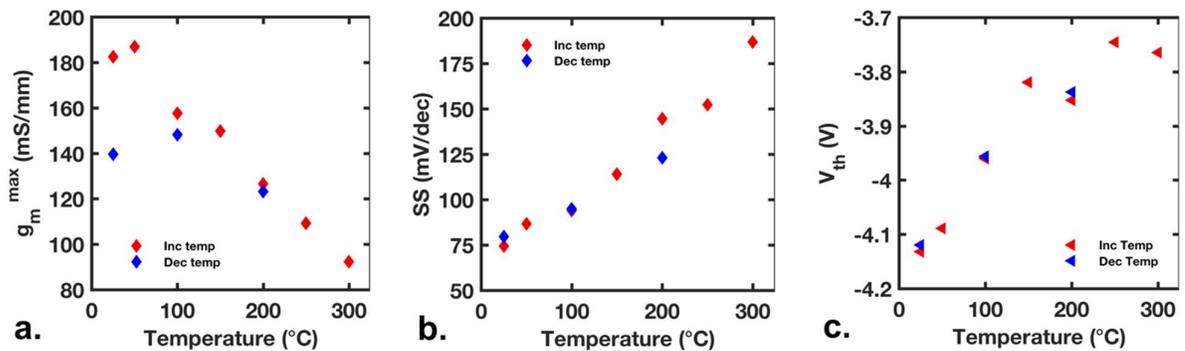

**Figure 4.** OFF-current (a), saturation current (b), ON/OFF ratio (c), vs. temperature.

The maximum transconductance ($g_m^{max}$), the subthreshold swing (SS), and threshold voltage ($V_{th}$) as a function of temperature are plotted in Figure 5. At 25°C, the SS is 75 mV/decade, and at 300°C, it is 187 mV/decade. These values are near the SS limits of 60 mV/decade and 110 mV/decade at 25°C and 300°C, respectively. The SS for high temperature operation is a critical parameter because it is expected to deteriorate linearly with temperature and represents the change in voltage required to transition from the ON-regime to the OFF-regime. The $g_m^{max}$ and the magnitude of $V_{th}$ both decrease with increasing temperature. The $g_m^{max}$ decreases with increasing temperature at a rate proportional to T$^{-0.97}$. Generally, $g_m^{max}$ is expected to deteriorate at a rate of $T^{-1.5}$, due to mobility degradation at the same rate. The slower degradation of transconductance with temperature is attributed to the decrease in the magnitude of $V_{th}$, which creates a competing effect.

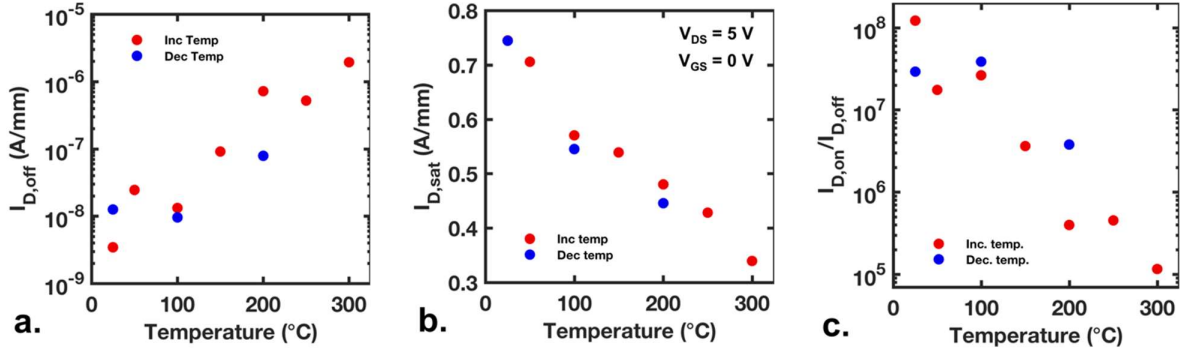

**Figure 5.** Max transconductance (a), subthreshold swing (b), threshold-voltage (c), vs. temperature.

To protect the $MoO_3$ gate, several passivation schemes were tested including atomic layer deposition (ALD) of $SiO_2$ and $Al_2O_3$. The $MoO_3$ gates were damaged by the precursors during ALD processing and no devices survived for testing. This illustrates the fragile nature of the $MoO_3$, and careful design of passivation schemes is needed to leverage the $MoO_3$ system at elevated temperatures.

Additionally, the $MoO_3$ was passivated with 50 nm PECVD silicon nitride (SiN) and survived. Testing the passivated $MoO_3$-gated HEMT up to 500°C the device survived. However, the leakage decreased even above the leakage of the as the deposited Mo-gated devices. The on-off ratios were $7.6\times10^3$ at RT, 110 at 300°C and, 15 at 500°C. It has been shown that SiN creates extra leakage paths at the passivation-III-nitride interface and through the barrier layer bulk, causing an increase in gate leakage current. [29, 30] This work initially aimed to utilize the same material platform previously demonstrated by the Maier paper by leveraging Mo as the gate and SiN for passivation on InAlN/GaN heterostructures. This work was done in air, a more extreme environment to the vacuum used in the Maier work. This work illustrates that when operating in air it is not enough to simply leverage Mo because $MoO_3$ volatilizes in the presence of air without passivation. However, SiN leakage paths must also be reduced for high ON/OFF ratio and high temperature operation.

**Conclusion**

High ON/OFF drain current ratios were found for d-mode $MoO_3$-gated InAlN/GaN HEMTs of $1.2 \times 10^8$ at 25°C and $1.2 \times 10^5$ at 300°C in air. Catastrophic device failure seen at 400°C in air is believed to be due to rapid volatilization of the oxidized molybdenum gate in the presence of water vapor. XPS results show that the electron-beam evaporated Mo is oxidized even without high temperature exposure after several weeks in air. It is confirmed with AES that only trace amounts of molybdenum remain on the surface of the InAlN layer after exposure to 400°C. The addition of SiN passivation enabled device operation up to 500°C but reduced the ON/OFF ratio by 5 orders of magnitude. While the results of $MoO_3$-gated InAlN/GaN HEMTs in air are promising for their low leakage and high ON/OFF ratio, the volatilization of the oxidized molybdenum in the presence of water vapor needs to be prevented through careful passivation choice.

**Acknowledgements**

This work was supported in part by the National Aeronautics and Space Agency through the High Operating Temperature Technology program under grant number NNX17AG44G. Part of this work was performed at the Stanford Nano Shared Facilities (SNSF), supported by the National Science Foundation under award ECCS-1542152. C. A. Chapin and S. R. Benbrook contributed equally to this work.